\documentstyle[psfig,sprocl]{article}

\def\Journal#1#2#3#4{{#1} {\bf #2}, #3 (#4)}

\def\MNRAS{\em MNRAS}
\newcommand{\la}{\mbox{$\ \stackrel{<}{_\sim}\ $}}

\newcommand{\hmompc}{\mbox{\,$h^{-1}$~Mpc}} 

\begin{document}

\title{LARGE GROUPS IN THE CHILE-UK QUASAR SURVEY}

\author{PETER R. NEWMAN, ROGER G. CLOWES}
\address{Centre for Astrophysics, University of Central Lancashire, \\
Preston PR1 2HE, U.K.}

\author{LUIS E. CAMPUSANO}

\address{Observatorio Astron\'{o}mico Cerro Cal\'{a}n, Universidad de
Chile, \\
Casilla 36D, Santiago, Chile}

\author{MATTHEW J. GRAHAM}
\address{Centre for Astrophysics, University of Central Lancashire, \\
Preston PR1 2HE, U.K.}

\maketitle

\abstracts{ 
The Chile-UK quasar survey, a new-generation 140~\mbox{deg$^{2}$}\ UVX
survey to $B = 20$, is now $\sim 25$\% complete.  The catalogue currently
contains 319 quasars and 93 emission line galaxies.  Using the
minimal-spanning tree method, we have independently confirmed the $\sim
200$\hmompc\ group of quasars at $z \simeq 1.3$ discovered by Clowes \&
Campusano (1991).  We have discovered a new $\sim 150$\hmompc\ group of 13
quasars at median $z \simeq 1.51$.  The null hypothesis of a uniform,
random distribution is rejected at a level of significance of $0.003$ for
both groups. 
}
  
\section{The Chile-UK quasar survey}

We are conducting an ultraviolet-excess (UVX) quasar survey using the
2\hbox{$^\circ$}-field, 128-fibre, multi-object spectrograph on the du~Pont 
2.5-m telescope at Las Campanas, to study large quasar groups
and the general clustering properties of quasars for $0.4 \la z \la 2.2$.
The survey includes the largest structure previously known: the $\sim
200$\hmompc\ group of $\ge 18$ quasars at $z \simeq 1.3$ (the CC91
group~\cite{clowesc91,clowescg95}).  Survey candidates are all objects
with $(U-B) \le -0.3$ and $16 \le B \le20$ on UK Schmidt Telescope plates 
digitized by SuperCOSMOS.  

\section{Minimal spanning tree identification of large groups}

\begin{figure}[t]
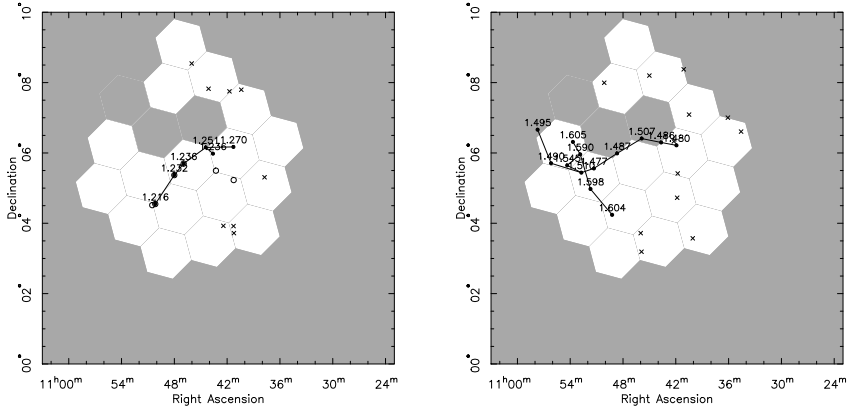


\parbox[c]{0.49\textwidth}{
  \psfig{file=fig_struc14_confproc.eps,angle=270,width=\hsize}}
\parbox[c]{0.49\textwidth}{
  \psfig{file=fig_struc19_confproc.eps,angle=270,width=\hsize}}

\caption{
The main survey area (shaded) and the area nominally observed so far
(unshaded -- the actual area observed is slightly larger).  Dots (labelled
with redshifts) show the survey quasars in MST groups A (left) and B
(right).  Lines show 2D projections of the 3D MST edges, each shorter than
the critical distance.  Crosses mark other UVX quasars in the same
redshift range but at larger separations. In the left panel, open circles
show the core MST group of 6 quasars from the CC91 objective-prism survey. 
Three quasars are common to the CC91 and UVX groups.  The other three CC91
quasars not in the UVX group are within the critical distance of the UVX
group, but are too faint to be in the UVX sample. 
}

\label{fig_groups}
\end{figure}

We used the MST method~\cite{grahamcc95} to identify large groups in the
quasar catalogue, and to determine their statistical significance.  The
critical separation distance which maximizes the number of groups of
connected quasars is 44\hmompc.  Figure \ref{fig_groups} shows the two
most significant groups.  Group A, 6 quasars at median $z \simeq 1.24$
coincides with the core of the CC91 group,\cite{graham97} independently
confirming this group.  Group B, 13 quasars at median $z \simeq 1.51$
spanning $\sim 150$\hmompc, is a new detection.  These results suggest
that parallel sheets or cellular structure and voids existed by $z \simeq
1.5$, and may be related to recent reports of large-scale structure from
observations of the galaxy distribution and of gas at high redshift. 

\section*{Acknowledgments}

We thank the Carnegie Institution of Washington and CTIO for telescope
time.  PRN is supported by a UK PPARC research studentship.  LEC was 
partially supported by FONDECYT grant 1970735.


\section*{References}
\begin{thebibliography}{99}


\bibitem{clowesc91}R.G. Clowes, L.E. Campusano,
\Journal{\MNRAS}{249}{218}{1991}. 

\bibitem{clowescg95}R.G. Clowes, L.E. Campusano, M.J. Graham in {\em 
Wide Field Spectroscopy and the Distant Universe}, eds. S.J. Maddox and 
A. Arag{\'{o}}n--Salamanca (World Scientific, Singapore, 1995), p. 400.

\bibitem{grahamcc95}M.J. Graham, R.G. Clowes, L.E. Campusano, 
\Journal{\MNRAS}{275}{790}{1995}.

\bibitem{graham97}M.J. Graham, PhD thesis, University of Central
Lancashire (1997).

\end{thebibliography}
\end{document}